\documentstyle[12pt]{article}
\begin{document}
\pagestyle{empty}
\parindent=0pt
\leftline{appeared in JMP October 1993.}
\vskip 10pt
\section*{TWISTOR PHASE SPACE DYNAMICS \break AND THE LORENTZ FORCE EQUATION.}
\vskip 10 pt
\leftline{By}
\leftline{Andreas Bette,}
\leftline{Stockholm University,}
\leftline{Department of Physics,}
\leftline{Box 6730,}
\leftline{S-113 85 STOCKHOLM,}
\leftline{SWEDEN.}
\vskip 10 pt
\leftline{fax +46-8347817 att.  Andreas Bette.}
\leftline{e-mail: $<$ab@vanosf.physto.se$>$.}
\vskip 50 pt
\centerline{ABSTRACT}

\vskip 15 pt

Using Lorentz force equation as an input a Hamiltonian mechanics on
the non-projective two twistor phase space TxT is formulated.

\vskip 10pt

Such a construction automatically reproduces dynamics of the intrinsic
classical relativistic spin.

\vskip 10pt

The charge appears as a dynamical variable.

\vskip 10pt

It is also shown that if the classical relativistic spin function on
TxT vanishes, the natural conformally invariant symplectic structure
on TxT reduces to the natural symplectic structure on the cotangent
bundle of the Ka{\l}u{\.z}a-Klein space.

\vfill
\eject

\section{INTRODUCTION.}

The classical motion of a relativistic electrically charged massive
and spinning particle exposed to an external
electromagnetic
field is, in Minkowski space, described by the Lorentz-Dirac (LD)
force equation and by the so called Bargmann, Michel, Telegdi (BMT)
equation for the intrinsic angular momentum (the spin).

If we by $X^{a}$, $P_{a}$, $S_{a}$, $F_{ab}$, $m^{2}:=P^{b}P_{b}$, $e$
and $g$ denote the four-position, the four-momentum, the Pauli-Luba{\'n}ski
four-vector, the external electromagnetic field tensor, the mass squared,
the charge and the gyromagnetic ratio of the particle then
these Poincar{\'e} covariant equations may be written as follows:

$$\dot X^{a} = P^{a},\eqno (1.1)$$
$$\dot P_{a} = eF_{ab}P^{b} + D_{a},\eqno (1.2)$$
$$\dot S_{a} = {ge\over 2}F_{ab}S^{b}+{ge\over 2m^{2}}(F_{ik}S^{i}P^{k})P_{a}
-{1 \over m^{2}}({\dot P_{k}}S^{k})P_{a}\eqno (1.3)$$

\vskip 10pt

where

$$P_{a}S^{a}=0 \eqno (1.4)$$

and

$$D_{a}P^{a}=0 \eqno (1.5).$$

\vskip 10pt

$D_{a}$ is a small space-like
correction four-vector (small compared with the space-like four-vector
$eF_{ab}P^{b}$) containing higher derivatives
of the external electromagnetic field $F_{kl}$, $F^{*}_{kl}$
and terms
nonlinear
in the spin variable $S^{i}$ [1,2].

\vskip 15pt

When the particle
forms (a classical limit of) an electron and the
radiation damping effects are neglected the value of $g$ equals $2$
(the Dirac value).

\vskip 15pt

The equations (1.1) - (1.5) are such that
the mass squared and the
spin squared of the particle:

$$m^{2}:=P_{a}P^{a} \ \ \ \ \ \ s^{2}:=-{1 \over m^{2}}S_{a}S^{a}\eqno (1.6)$$

\vskip 10pt

are constants of the motion.

\vskip 15pt

The dot in (1.1) - (1.3) denotes differentiation with respect
to a real parameter $l$
which is, by virtue of (1.1), linearly related to the proper
time ${\tau}$ of the particle by:

$${\tau}= {\pm}ml+{\tau_{0}}\eqno (1.7).$$

\vskip 10pt

${\tau_{0}}$ is an arbitrary real number representing the freedom
of choice of the origin of the proper time.

\vskip 15pt

$F_{ab}$ denotes the value of the
external electromagnetic field evaluated at the particle's four-position
$X^{a}$.
Consequently, the four-position coincides with
the location of the
charge $e$.

\vskip 15pt

In this paper we assume that $D_{a}=0$ in (1.2) and then
examine (1.1) and (1.2)
using two distinct twistors as variables.

\vskip 15pt




This analysis
will automatically
produce the BMT equation in (1.3) with $g=2$ [4].





\vskip 15pt

In the next section we give a physical interpretation to
the sixteen variables corresponding to a point in the space
of two twistors TxT [3,4].

\vskip 15pt

In section three the free particle
symplectic potential on TxT is expressed using these physical variables. The
non-uniqeness
of choice of the free particle Hamiltonian is disscused.

\vskip 10pt

The free particle equations of motion given as a canonical flow in the
phase space of two twistors are presented in twistors' Weyl spinor coordinates
as well as in Poincar{\'e} covariant physically interpretable coordinates.
These has been presented before [3] in a somewhat preliminary shape.

\vskip 15pt

In section four a deformed Poincar{\'e}
covariant symplectic structure and a deformed Poincar{\'e} scalar
Hamiltonian function on TxT are presented.
The new Poincar{\'e} covariant
flow in TxT
canonical with respect to the deformed symplectic structure
and generated by the deformed Hamiltonian
reproduces (1.1) - (1.4) (with $D_{a}=0$ and $g=2$)
and also produces certain additional equations of motion.
The latter arise because TxT is sixteen dimensional while the number
of independent variables describing the particle
according to (1.1) - (1.4) is only twelve (the four-position,
the four-momentum, the Pauli-Luba{\'n}ski spin four-vector
fulfilling (1.4) and the charge).

Our attempt to interpret physically the remaining
four variables is presented already in section two.
However, a (partial) confirmation of the correctness of
these tentative identifications
is provided first
when the interaction with an external electromagnetic
field
is "switched" on. This is done in section four.

\vskip 15pt

A first version of the material contained in section four
appeared in [5] where the electric charge was not defined as a
dynamical variable. This weakness of the model is removed in section
four of the present paper.

\vskip 15pt

In the appendix the formal proof of the statements made in section four
is presented.

\vskip 15pt

Upper case latin letters with lower case greek indices denote twistors.

Upper case latin letters with lower case latin indices denote four-vectors
and four-tensors.

Lower case greek letters with upper case latin indices (either primed
or unprimed) denote Weyl spinors.

The Minkowski metric has the signature $+---$.

\section {PHYSICAL VARIABLES IDENTIFIED \break AS FUNCTIONS ON T$\Delta$T.}

The symbol TxT usually denotes the direct product of two twistor spaces.
However, in our investigations, we will not be using the whole of TxT
but rather T$\Delta$T which from now on will denote the space TxT
less its diagonal i.e.
T$\Delta$T := $\{$TxT - $\{$(t, t) $\epsilon$ TxT; t $\epsilon$ T$\} \}$.

\vskip 15pt

The twistor coordinates of a point in T$\Delta$T will be expressed in
terms of two Weyl spinors:

$$Z^{\alpha} = (\omega^{A},\ \pi_{A^\prime})
\qquad and \qquad W^{\alpha} = (\lambda^{A},\ \eta_{A^\prime})\eqno (2.1)$$

\vskip 10pt

or dually (complex conjugation):

$$\overline Z_{\alpha} = ( \overline \pi_{A},\ \overline \omega^{A^\prime})
\qquad and \qquad \overline W_{\alpha} =
(\overline \eta_{A},\ \overline \lambda^{A^\prime})\eqno(2.2).$$

\vskip 10pt

Using these two twistors and their twistor conjugates four
independent  conformally (SU(2,2))
scalar functions may be formed on T$\Delta$T [4,7,8]:

$$s_{1} = Z^{\alpha}\overline Z_{\alpha} \qquad and \qquad
s_{2} = W^{\alpha}\overline W_{\alpha}\eqno (2.3)$$

$$a = Z^{\alpha}\overline W_{\alpha} \qquad and \qquad
\overline a = W^{\alpha}\overline Z_{\alpha}\eqno (2.4).$$

\vskip 10pt

In addition, the following two Poincar\'e scalar functions
may also be defined on T$\Delta$T:

$$f = \pi^{A^\prime}\eta_{A^\prime} \qquad and \qquad
{\overline f} = \overline \pi^{A}\overline \eta_{A}\eqno (2.5).$$

\vskip 10pt

The scalar functions introduced above may be represented by six real
valued functions on T$\Delta$T given by:

$$e =  s_{1} + s_{2} \qquad and \qquad k =  s_{1} - s_{2} \eqno (2.6)$$

$$\mid a\mid \qquad and \qquad \vartheta = arga = -arg{\overline a}\eqno
(2.7)$$

$$\mid f\mid \qquad and \qquad \varphi = argf = -arg{\overline f}\eqno (2.8).$$

\vskip 15pt

Below, Poincar{\'e} covariant
functions
on T$\Delta$T  will be identified as
physical quantities
according to the following recipe [4,5]
(we employ here the abstract index notation according to Penrose [6]):

$$P_{a} := \pi_{A^\prime}{\overline\pi}_A +
\eta_{A^\prime}{\overline\eta}_A\eqno (2.9)$$

\vskip 10pt

will denote a massive particle's four-momentum
expressed as a sum of
the four-momenta of its two massless parts.

\vskip 15pt

$$X^{a} := {1\over 2}(Z^{a} + {\overline Z}^{a})\qquad where \qquad
Z^{a} := {i\over f}(\omega^{A}\eta^{A^\prime}-\lambda^{A}\pi^{A^\prime})
\eqno (2.10)$$

\vskip 10pt

will denote a massive particle's four-position in Minkowski space time.

\vskip 15pt

A massive particle's Pauli-Luba{\'n}ski spin four-vector will be given
by:

$$S_{a} := {k\over 2}(\pi_{A^\prime}{\overline\pi}_A - \eta_{A^\prime}
{\overline\eta}_A) + a\eta_{A^\prime}{\overline\pi}_A +
{\overline a}\pi_{A^\prime}{\overline\eta}_A\eqno (2.11).$$

\vskip 10pt

The definition in (2.11)
is dictated by the assumption that
a massive particle's  four angular
momentum should be a sum of the four angular
momenta of its two massless parts (see e.g. [3]).

\vskip 15pt

Note that $P_{a}$ and $S_{a}$ are automatically orthogonal to each other
i.e. we always have $P_{a}S^{a}=0$.

\vskip 15 pt

{}From the above it follows that the imaginary part of $Z^{a}$:

$$Y^{a} = {1\over 2i}(Z^{a} - {\overline Z}^{a})=\eqno (2.12)$$

\vskip 10pt

may be written as:

$$= {1\over 2f\overline f}\Big[(a\eta^{A^\prime}{\overline\pi}^A +
{\overline a}\pi^{A^\prime}{\overline\eta}^A) -
s_{1}\eta^{A^\prime}{\overline\eta}^A -
s_{2}\pi^{A^\prime}{\overline\pi}^A\Big] =\eqno (2.13)$$

\vskip 10pt

or as:

$$={1\over 2f\overline f}(S^{a} - {e\over 2} P^{a})\eqno (2.14).$$

\vskip 10pt

{}From the definitions above it also follows that on T$\Delta$T the mass
function of the particle is given by

$$m={\sqrt 2} \mid f\mid \eqno (2.15)$$

\vskip 10pt

while its spinfunction by:

$$s=\sqrt {{1\over 4} k^{2} + {\mid a\mid}^{2}}\eqno (2.16).$$

\vskip 10pt

A space-like plane spanned by two mutually orthogonal unit four-vector
valued functions on T$\Delta$T orthogonal to $S_{a}$ and $P_{a}$:

$$E_{a} := {i \over (m {\mid a \mid})}
(a\eta_{A^\prime}{\overline\pi}_A -
{\overline a}
\pi_{A^\prime}{\overline\eta}_A)
\eqno (2.17).$$

$$F_{a} := {1 \over (sm {\mid a \mid})}
[{k \over 2}(a\eta_{A^\prime}{\overline\pi}_A +
{\bar a}
\pi_{A^\prime}{\overline\eta}_A) -
{\overline a}a
(\pi_{A^\prime}{\overline\pi}_A - \eta_{A^\prime}
{\overline\eta}_A)]
\eqno (2.18)$$

\vskip 10pt

may be thought of as a polarization plane rigidly
attached to the massive particle at its four-position
$X^{a}$ in Minkowski space [3,4].

\vskip 15pt

In effect, all four four-vectors
${P_{a}/m}$, ${S_{a}/(sm)}$,
$E_{a}$ and $F_{a}$ span an orthonormal tetrad rigidly attached
to the particle at its four-position
$X^{a}$ in Minkowski space. The number of variables represented by the
functions defining this tetrad
is six, the number of variables represented by
the scalar functions is also six, while the four-position represents
four variables; sixteen variables altogether.

\vskip 15pt

With these identifications the
inverse relations expressing twistor coordinates in (2.1) - (2.2)
as functions of the introduced Poincar{\'e} covariant physical
variables and the scalars in (2.6) - (2.8) (note that according
to (2.15) and (2.16) two of these scalars have a clear physical
interpretation) are almost immediate.

\vskip 15pt

The spinor $\pi_{A^\prime}$ up to its phase is given by:

$$\pi_{A^\prime}{\overline\pi}_A ={{1 \over 2}(P_{a} + {k \over 2s^{2}}S_{a} -
{m {\mid a \mid} \over s}F_{a})} \eqno (2.19),$$

\vskip 10pt

the spinor $\eta_{A^\prime}$ up to its phase is given by:

$$\eta_{A^\prime}{\overline\eta}_A = {{1 \over 2}(P_{a} -
{k \over 2s^{2}}S_{a} +
{m {\mid a \mid} \over s}F_{a})} \eqno (2.20),$$

\vskip 10pt

while the phase $\alpha$ of the spinor $\pi_{A^\prime}$ is given by:

$$\alpha={1 \over 2}(argf + arg a) = {1 \over 2}(\varphi + \vartheta)
\eqno (2.21),$$

\vskip 10pt

and the phase $\beta$ of the spinor $\eta_{A^\prime}$ by:

$$\beta = {1 \over 2}(argf - arg a) = {1 \over 2}(\varphi - \vartheta)
\eqno (2.22).$$

\vskip 10pt

The relations in (2.21) and (2.22) follow from (2.5) and from
the fact that the conformal
complex valued scalar $a$ in (2.4) may be written as:

$$a=-2Y^{A{A^\prime}}{\overline \eta}_{A}{\pi_{A^\prime}} \eqno (2.23)$$

\vskip 10pt
where $Y^{a}$ is a real four-vector valued function on T$\Delta$T introduced in
(2.12).

\vskip 15pt

The remaining spinors are given by (see (2.14) and (2.15)):

$$\omega^{A} = iX^{AA^{\prime}}\pi_{A^\prime} -
{1 \over m^{2}}(S^{AA^{\prime}}\pi_{A^\prime} -
{e \over 2} P^{AA^{\prime}}\pi_{A^\prime})
\eqno (2.24)$$

and

$$\lambda^{A} = iX^{AA^{\prime}}\eta_{A^\prime}
-{1 \over m^{2}}(S^{AA^{\prime}}\eta_{A^\prime} -
{e \over 2} P^{AA^{\prime}}\eta_{A^\prime}) \eqno (2.25).$$

\section {THE FREE PARTICLE MOTION.}

The two twistor space T$\Delta$T possesses a natural
(free particle) symplectic structure given by [7,8]:

$$\Omega_{0} = i(dZ^{\alpha}\wedge d\overline Z_{\alpha} + dW^{\alpha}\wedge
d\overline W_{\alpha}) \eqno (3.1).$$

\vskip 10pt

$\Omega_{0}$ may be regarded as exterior derivative of a
one-form $\gamma_{0} \ (\Omega_{0} = d\gamma_{0})$ given by:

$$\gamma_{0} = {i\over 2}(Z^{\alpha}d\overline Z_{\alpha} -
\overline Z_{\alpha}dZ^{\alpha} + W^{\alpha}d\overline W_{\alpha} -
\overline W_{\alpha}dW^{\alpha})\eqno (3.2).$$

\vskip 10pt

Using the introduced Poincar{\'e} covariant physical functions on T$\Delta$T,
$\gamma_{0}$ may also
be written as:

$$\gamma_{0} = P_{j}dX^{j} + {1\over 2}ed\varphi - {1\over 2}kd\vartheta +
({k^{2} \over 4s}F_{j} +
{{\mid a \mid}k \over 2ms^{2}}S_{j} +
{{\mid a \mid} \over m}P_{j})dE^{j}
\eqno (3.3)$$

\vskip 10pt

or equivalently

$$\gamma_{0} = P_{j}dX^{j} + {1\over 2}ed\varphi - {1\over
2}kd\vartheta + {k \over 2m}(iM_{j}d{\bar M}^{j} - i{\bar
M}_{j}dM^{j}) + $$

$$+{i{\bar a} \over m^{2}}M_{j}dP^{j}- {ia \over
m^{2}}{\bar M}_{j}dP^{j} \eqno (3.4)$$

\vskip 10pt

where $M_{j}$ is a complex null four-vector valued function on T$\Delta$T
given by:

$$M_{a}:=\pi_{A^{\prime}}{\bar \eta}_{A} \eqno (3.4a).$$


\vskip 10pt

{}From (3.3) or (3.4) we notice a remarkable fact that for $a=k=0$
i.e. for the vanishing value of the spin function on T$\Delta$T, the
conformally invariant symplectic potential $\gamma_{0}$ in (3.2)
(and thereby also the symplectic structure $\Omega_{0}$ in (3.1))
reduces to the natural symplectic potential
(while $\Omega_{0}$ reduces to the natural symplectic structure)
on the cotangent bundle of the Ka{\l}u{\.z}a-Klein space.
This suggests that $e$ should be identified with the electric
charge of the particle.

\vskip 15pt

To generate the free motion of a massive particle built up of
the two twistors we used in [3] a Hamiltonian:

$$H_{0_1} = m^{2} + s^{2}\eqno (3.5)$$

\vskip 10pt

and a somewhat modified one in [4]:

$$H_{0_2} = {1\over 2}(m^{2} + s^{2}) \eqno (3.6).$$

\vskip 10pt

Any such a change is of no importance as long as
$H_{0}$ is a function on T$\Delta$T  such that:

$$H_{0} = H_{0}(m,\ s) \eqno (3.7).$$

\vskip 10pt

The flow will always correspond to a free particle motion in Minkowski space.
In fact
any function such that:

$$H_{0} = H_{0}(m,\ s,\ k,\ e)\eqno (3.8)$$

\vskip 10 pt

describes a free particle. As $m$, $s$, $k$ and $e$ are mutually
(Poisson) commuting functions the different choices of $H_{0}$
may correspond to different motions of the internal physical variables
represented by
$\varphi$ and
$\vartheta$.

\vskip 10pt

But in most cases different choices of $H_{0}$ simply correspond to a
reparametri- zation of the canonical flow lines.

\vskip 15pt

At this non-quantum level there is thus quite a large freedom
of  choice of the free particle
Hamiltonian $H_{0}$. On the quantum level of this approach
one should, on the other hand, expect essential differences
depending on the choice of ${\hat H}_{0}$.

\vskip 10pt

In this paper, for simplicity, we choose $H_{0}$ as:

$$H_{0}:={1 \over 2}m^{2} + (s^{2} - {1\over 4}e^{2}) \eqno (3.9)$$

\vskip 10pt

which written out in terms of the introduced scalar functions yields:

$$H_{0} := f{\overline f} + {1\over 4}k^{2} + a{\overline a} - {1\over 4}e^{2}
= f{\overline f} - s_{1}s_{2} + a{\overline a} \eqno (3.10).$$

\vskip 10pt

The chosen $H_{0}$ and $\Omega_{0}$ in (3.1) generate the following equations
of motion in T$\Delta$T:

$$\dot\omega^A = -if\overline\eta^A + ia\lambda^A - is_{2}\omega^A
\eqno (3.11)$$

$$\dot\pi_{A^\prime} = ia\eta_{A^\prime} - is_{2}\pi_{A^\prime} \eqno (3.12)$$

$$\dot\lambda^A = if\overline\pi^A + i{\overline a}\omega^A -
is_{1}\lambda^A \eqno (3.13)$$

$$\dot\eta_{A^\prime} = i{\overline a}\pi_{A^\prime} - is_{1}\eta_{A^\prime}
\eqno (3.14)$$

\vskip 10pt

and their complex conjugates (c.c.).

\vskip 15pt

The above equations, written out using functions representing
the physical variables
as previously identified, read:

$${\dot e}= 0 \qquad and \qquad {\dot k} = 0 \eqno (3.15)$$

$${\dot \varphi} = - e \qquad and \qquad {\dot \vartheta} = 0 \eqno (3.16)$$

$${\dot X}^{a} = P^{a} \eqno (3.17)$$

$${\dot P}_{a} = 0 \qquad and \qquad {\dot S}_{a} = 0  \eqno (3.18)$$

$${\dot E}_{a} = 2s F_{a} \eqno (3.19)$$

$${\dot F}_{a} = -2s E_{a} \eqno (3.20).$$

\vskip 10pt

{}From (3.19), (3.20) and (1.7) it follows that, with our choice of $H_{0}$,
the introduced polarization plane rigidly
attached to the particle rotates with an angular velocity equal
to $(2s/m)$ [3].






\section {MOTION IN AN EXTERNAL ELECTRO \break MAGNETIC FIELD.}

In this section we identify the function $e$ on T$\Delta$T with
the electric charge of the particle. The deformed Poincar{\'e}
covariant symplectic potential on T$\Delta$T we define as:

$$\gamma = \gamma_{0}+eA_{j}dX^{j} \eqno (4.1)$$

\vskip 10pt

where $X^{a}$ is a four vector-valued function on T$\Delta$T given by (2.10)
and
where $A_{j}=A_{j}(X^{a})$
denotes an external electromagnetic four-potential. $A_{j}$ is in this way
a four-vector valued function defined on T$\Delta$T.
$\gamma_{0}$ is given by (3.3) (or equivalently
by (3.2) or (3.4)).

\vskip 15pt

The external derivative of $\gamma$ gives us the deformed symplectic
structure on T$\Delta$T:

$$\Omega  = \Omega_{0} + de\wedge dX^{j}A_{j}
+ {1\over 2}eF_{jk}dX^{j}\wedge dX^{k} \eqno (4.2)$$

\vskip 10pt

where $F_{jk}=F_{jk}(X^{a})$ denotes the
electromagnetic field tensor formed from
$A_{j}$. $\Omega_{0}=d\gamma_{0}$.

\vskip 10pt

Note that for $a=k=0$, $\gamma$ and thereby $\Omega$ may be regarded
as a deformation of the
natural symplectic potential and natural symplectic structure on
the cotangent bundle of the Ka{\l}u{\.z}a-Klein space.

\vskip 15pt

As the deformed Hamiltonian function on T$\Delta$T we take:

$$H = H_{0} + {e \over m^{2}}F^{*}_{jk}S^{j}P^{k} \eqno (4.3)$$

\vskip 10pt

where $F^{*}_{jk}=F^{*}_{jk}(X^{a})$
represents on T$\Delta$T the dual of the external
electromagnetic tensor field.

\vskip 15pt

It is shown in the appendix that, with respect to $\Omega$, $H$
generates a Poincar{\'e} covariant canonical flow
in T$\Delta$T provided Maxwell's empty space
equations are fulfilled at the location of the particle:

$$F^{*}_{\lbrack jk,n\rbrack} = 0 \eqno (4.4).$$

\vskip 10pt

For future reference we note that
using (2.12), (2.14), (2.15) and the skew symmetry of the dual of the
external electro-magnetic field tensor the generating function $H$ may
also be written as:



$$H = H_{0} + eF^{*}_{ik}Y^{i}P^{k} \eqno (4.5).$$

\vskip 10pt

Expressed in twistor coordinates the flow canonical with respect to
$\Omega$ and generated by $H$ is given by the following equations of
motion (see proof in the appendix):

$$\dot\omega^A = -if\overline\eta^A + ia\lambda^A - is_{2}\omega^A+$$

$$ + e\mu^A_{\ B}Y^{B{B^\prime}}\pi_{B^\prime} +
ieX^{AA^\prime}{{\overline \mu}_{A^\prime}\ ^{B^{\prime}}}\pi_{B^\prime}
+ iC\omega^A \eqno (4.6)$$

$$\dot\pi_{A^\prime} = ia\eta_{A^\prime} - is_{2}\pi_{A^\prime} +
e{{\overline \mu}_{A^\prime}\ ^{B^{\prime}}}\pi_{B^\prime} +
iC\pi_{A^\prime} \eqno (4.7)$$

$$\dot\lambda^A = if\overline\pi^A + i{\overline a}\omega^A - is_{1}\lambda^A
+ $$

$$+e\mu^A_{\ B}Y^{B{B^\prime}}\eta_{B^\prime} +
ieX^{AA^\prime}{{\overline \mu}_{A^\prime}\ ^{B^{\prime}}}\eta_{B^\prime} +
iC\lambda^A \eqno (4.8)$$

$$\dot\eta_{A^\prime} = i{\overline a}\pi_{A^\prime} - is_{1}\eta_{A^\prime} +
e{{\overline \mu}_{A^\prime}\ ^{B^{\prime}}}\eta_{B^\prime} + iC\eta_{A^\prime}
\eqno (4.9)$$

where

$$C = (F^{*}_{ik}Y^{i}P^{k}-A^{i}P_{i}) \eqno (4.10)$$

\vskip 10pt

and where $\mu_{AB} = \mu_{AB}(X^{c})$ is a spinor field corresponding
to $F_{ab}=F_{ab}(X^{c})$ [6]:

$$\mu_{AB}={1 \over 2}F_{AA^{\prime}B}\ ^{A^{\prime}} \eqno (4.11).$$

\vskip 10pt

Conversely one has [6]:

$$F_{ab} = \mu_{AB}\epsilon_{{A^\prime}{B^\prime}} +
{\overline \mu}_{{A^\prime}{B^\prime}}\epsilon_{AB} \eqno (4.12)$$

$$F^{*}_{ab} = i{\overline \mu}_{{A^\prime}{B^\prime}}\epsilon_{AB} -
i\mu_{AB}\epsilon_{{A^\prime}{B^\prime}}  \eqno (4.13).$$

\vskip 10pt

Written out in terms of the introduced Poincar{\'e} covariant physical
variables the above equations of motion read:

$$\dot X^{j} = P^{j} \qquad and \qquad \dot P_{j} = eF_{jk}P^{k} \eqno (4.14)$$

$$\dot S_{j} = eF_{jk}S^{k} \eqno (4.15)$$

$$\dot e=0 \qquad and \qquad \dot k=0 \eqno (4.16)$$

$$\dot \vartheta = 0 \eqno (4.17)$$

$$\dot \varphi = -e - 2P_{j}A^{j} + {2 \over m^{2}}F^{*}_{jk}S^{j}P^{k}
\eqno (4.18)$$

$${\dot E}_{j} = 2s F_{j} + e F_{kj}E^{k} \eqno (4.19)$$

$${\dot F}_{j} = -2s E_{j} + e F_{kj}F^{k} \eqno (4.20).$$

\vskip 15pt

As may be seen the equations in (4.14) and (4.15) are the same as
those in (1.1) - (1.3) (with $D_{j}=0$ and $g=2$) while the relation
in (1.4) is automatically fulfilled because of the way $S_{j}$ and
$P_{j}$ were defined in (2.9) and (2.11).

\vskip 15pt

The charge function $e$ appears as a dynamical variable and according to
(4.16) is a constant of motion.

\vskip 15pt

Conformally scalar functions $k$ and $\vartheta$
in (4.16) and (4.17) do not yet have any
clear physical interpretation. They form two ((Poisson) non-commuting)
constants of motion.

\vskip 15pt

The first two terms in (4.18) correspond to the Aharonov-Bohm effect
while the third term arises because of the non-vanishing intrinsic
spin of the particle.

\vskip 15pt

The motion of the polarization plane is given by (4.19) and (4.20).

\vskip 15 pt

Finally we note that the equations of motion in (4.6) - (4.9) may also
be written in a twistor covariant way i.e.  entirely in terms of
$Z^{\alpha}$ and $W^{\alpha}$:

$${\dot Z}^{\alpha} = (if + l_{1}f)
I^{\alpha \beta}{\bar W}_{\beta}
+(ia - {\bar c}_{2})W^{\alpha}  - (is_{2} + iC + {\bar c}_{3})Z^{\alpha}
- b I^{\alpha \beta}{\bar Z}_{\beta} \eqno (4.21)$$

$${\dot W}^{\alpha} = (if + l_{2}f)I^{\beta \alpha}{\bar Z}_{\beta}
+(i{\bar a} + {\bar c}_{1})Z^{\alpha}  - (is_{1} + iC - {\bar c}_{3})W^{\alpha}
- {\bar b} I^{\beta \alpha}{\bar W}_{\beta} \eqno (4.22)$$

\vskip 10pt

where $I^{\alpha \beta}$ is the so called infinity twistor and where
$c_{1}$, $c_{2}$, $c_{3}$ are certain, conveniently chosen, complex
valued Poincar{\'e} scalar functions on T$\Delta$T describing the
external electromagnetic field ($e$ and ${\bar f}$ are defined in
(2.5) - (2.6)):

$$c_{1}={e{\mu^{AB}}{\bar \eta}_{A}{\bar \eta}_{B} \over {\bar f}}
\eqno (4.23)$$

$$c_{2}={e{\mu^{AB}}{\bar \pi}_{A}{\bar \pi}_{B} \over {\bar f}} \eqno (4.24)$$

$$c_{3}=-{e{\mu^{AB}}{\bar \pi}_{A}{\bar \eta}_{B} \over {\bar f}}
 \eqno (4.25).$$

\vskip 15pt

In (4.21) - (4.22) $l_{1}$ and $l_{2}$ are real valued Poincar{\'e}
scalar functions on T$\Delta$T while b is a complex valued
Poincar{\'e} scalar function on T$\Delta$T given by (see (2.3) and
(2.4)):

$$l_{1}=-{1 \over m^{2}}[ac_{2} + {\bar a}{\bar c}_{2} + s_{1} (c_{3}
 + {\bar c}_{3})] \eqno (4.26),$$

$$l_{2}={1 \over m^{2}}[{\bar a}c_{1} + a {\bar c}_{1} + s_{2} (c_{3}
 + {\bar c}_{3})] \eqno (4.27),$$

$$b={1 \over m^{2}}[a({\bar c}_{3}-c_{3}) -  ({\bar c}_{2}s_{2} +
c_{1}s_{1})] \eqno (4.28).$$

\vskip 15pt

The deformed symplectic potential in (4.1), the corresponding
symplectic structure in (4.2), the deformed Hamiltonian in (4.3) (or
(4.5)) may all be written in a twistor covariant way i.e. entirely
in terms of $Z^{\alpha}$ and $W^{\alpha}$ and the infinity twistor
$I^{\alpha \beta}$.
The arising expressions are, however,  quite complicated and not especially
illuminating from the physical point of view. In this paper we
therefore omit their presentation.

\section {CONCLUSIONS AND REMARKS.}

In this paper we describe the dynamics of a relativistic charged
particle with spin in an external electro-magnetic field using
two-twistor phase space T$\Delta$T. We have shown that there exists
a Hamiltonian dynamics on T$\Delta$T which after passing
to space-time coordinates reproduces the Lorentz force dynamics
and Bargmann-Michel-Telegdi
dynamics (with $g=2$) and also indicates connection to the
Ka{\l}u{\.z}a-Klein space dynamics. Conversely, one can say that
there exists a sort of the square root of the Lorentz force dynamics
which is realized as a Hamiltonian dynamics on T$\Delta$T.

\vskip 15pt

It will be interesting to see how the quantized version of the above
formalism corresponds to the Dirac equation coupled to an external
electromagnetic field. It seems that the approach developed by A.
Odzijewicz and his group [9-12] would be of great value here.

\section {REFERENCES.}

\leftline{[1] F. Rohrlich, "Classical Charged Particles", Addison-Wesley, 1965,
sect. 7-4,}

\vskip 10pt

\leftline{[2] L.D. Landau and E.M. Lifshitz, "The Classical Theory of Fields",}
 \leftline {Pergamon Press Ltd., 1985, sect. 76,}

\vskip 10pt

\leftline{[3] A. Bette, J. Math. Phys., Vol. 25, No. 8, 2456-2460, August
1984,}

\vskip 10pt

\leftline{[4] A. Bette, Rep. Math. Phys., Vol. 28, No. 1, 133-140, 1989,}

\vskip 10pt

\leftline{[5] A. Bette, J. Math. Phys., Vol. 33, No. 6, 2158-2163, June 1992,}

\vskip 10pt

\leftline{[6] R. Penrose and W. Rindler, "Spinors and Space-Time"-(Cambridge}
\leftline{monographs on mathematical physics) Vol. 1: "Two-spinor calculus and}
\leftline{relativistic fields 1. Spinor analysis", Cambridge University Press,
1984,}

\vskip 10pt

\leftline{[7] K.P. Tod, Massive Spinning Particles and Twistor Theory, Doctoral
}
\leftline{Dissertation, Mathematical Institute, University of Oxford, Oxford,
1975,}

\vskip 10pt

\leftline{[8] K.P. Tod, Rep. Math. Phys., Vol. 7, No. 3, 339-346, 1977,}

\vskip 10pt

\leftline{[9] A. Odzijewicz, Comm. Math. Phys., Vol. 107, 561-575,
1986,}

\vskip 10pt

\leftline{[10] A. Karpio, A.Krysze{\'n}, A. Odzijewicz, Rep. Math. Phys.,}
\leftline{Vol. 24, No. 1, 65-80, 1986,}

\vskip 10pt

\leftline{[11] A. Odzijewicz, Comm. Math. Phys., Vol. 114, 577-597,
1988,}

\vskip 10pt

\leftline{[12] A. Odzijewicz, Comm. Math. Phys., Vol. 150, 385-413, 1992.}

\vskip 25pt

\section {APPENDIX; A FORMAL PROOF OF \break THE MAIN RESULT OF SECT. 4}

In order to prove that the Hamiltonian in (4.5) and the symplectic structure
$\Omega$ in (4.2) generate equations (4.6) - (4.9) which, in turn, imply
(4.14) - (4.20) we have to prove that:

$$V{_{_{\_\!\_\!\_\!\_\!}}{\!|}}\Omega=-dH  \eqno (A.1)$$

\vskip 10pt

where $H$ is that in (4.5) $\Omega$ is that in (4.2) and where

$$ V = V_{0}+ V_{1} \eqno (A.2)$$

\vskip 10pt

where the vector-field $V_{0}$ according to (3.12)-(3.15) is given by:

$$V_{0}=(-if\overline\eta^A + ia\lambda^A - is_{2}\omega^A)
{\partial \over {\partial \omega^A}}
+ (ia\eta_{A^\prime} - is_{2}\pi_{A^\prime}){\partial \over {\partial
\pi_{A^\prime}}}+ $$

$$+(if\overline\pi^A + i{\overline a}\omega^A -
is_{1}\lambda^A){\partial \over {\partial \lambda^A}} +
(i{\overline a}\pi_{A^\prime} - is_{1}\eta_{A^\prime})
{\partial \over {\partial \eta_{A^\prime}}} + c.c. \eqno (A.3),$$

\vskip 10pt

or using the introduced four-vector variables (see (3.16)-(3.21)):

$$V_{0}=P^{j}{\partial \over {\partial X^{j}}} -
e {\partial \over {\partial \varphi}}
-2sE^{j}{\partial \over {\partial F^{j}}}
 + 2sF^{j}{\partial \over {\partial E^{j}}} \eqno (A.3a).$$

\vskip 10pt

The vector field $V_{1}$ is according to (4.6)-(4.9) given by:

$$V_{1}= (e\mu^A_{\ B}Y^{B{B^\prime}}\pi_{B^\prime} +
ieX^{AA^\prime}\overline \mu_{A^\prime}\ ^{B^{\prime}}\pi_{B^\prime} +
iC\omega^A) {\partial \over {\partial \omega^A}}+ c.c.+$$
$$+(e\overline\mu_{A^\prime}\ ^{B^{\prime}}\pi_{B^\prime} +
iC\pi_{A^\prime}){\partial \over {\partial \pi_{A^\prime}}}+ c.c.+$$

$$+(e\mu^A_{\ B}Y^{B{B^\prime}}\eta_{B^\prime} + ieX^{AA^\prime}
\overline\mu_{A^\prime}\ ^{B^{\prime}}\eta_{B^\prime} + iC\lambda^A)
{\partial \over {\partial \lambda^A}} + c.c.+$$
$$+(e{\overline\mu}_{A^\prime}\ ^{B^{\prime}}\eta_{B^\prime} +
iC\eta_{A^\prime}){\partial \over {\partial \eta_{A^\prime}}}+ c.c.
\eqno (A.4).$$

or using the introduced four-vector variables (see (4.14)-(4.20)):

$$V_{1}=eF_{jk}P^{k}{\partial \over {\partial P^{j}}} +
2C {\partial \over {\partial \varphi}} +
F^{kj}F_{k}{\partial \over {\partial F^{j}}}+
F^{kj}E_{k}{\partial \over {\partial E^{j}}} \eqno (A.4a).$$

\vskip 10pt

To facilitate the caculations
the inner product on the left hand side of (A.1) may be split
into a sum of partial inner products:

$$V{_{_{\_\!\_\!\_\!\_\!}}{\!|}}\Omega=V_{0}{_{_{\_\!\_\!\_\!\_\!}}{\!|}}\Omega_{0}
+V_{0}{_{_{\_\!\_\!\_\!\_\!}}{\!|}}\Omega_{1}+V_{1}{_{_{\_\!\_\!\_\!\_\!}}{\!|}}
\Omega_{0}+V_{1}{_{_{\_\!\_\!\_\!\_\!}}{\!|}}\Omega_{1} \eqno (A.5)$$

\vskip 10pt

where

$$\Omega = \Omega_{0} + \Omega_{1} \eqno (A.6)$$

$$\Omega_{0} =  i(dZ^{\alpha}\wedge d\overline Z_{\alpha} +
dW^{\alpha}\wedge d\overline W_{\alpha}) \eqno (A.7)$$
and
$$\Omega_{1} = de\wedge dX^{i}A_{i} + {1\over 2}eF_{ik}dX^{i}\wedge dX^{k}
\eqno (A.8).$$

\vskip 10pt

By assumption, which may be checked by direct calculations, one has:

$$V_{0}{_{_{\_\!\_\!\_\!\_\!}}{\!|}}\Omega_{0}=
-d({1\over 2}m^{2} + s^{2} - {1 \over 4}e^{2})
\eqno (A.9).$$

\vskip 10pt

Using the fact (see (3.16)-(3.21)) that:

$$V_{0}=P^{i}{\partial \over {\partial X^{i}}}+ \eqno (A.10)$$

\vskip 10pt

\centerline{ + terms in directions linearly independent of
${\partial \over {\partial X^{i}}}$}

\vskip 15 pt

and that $V_{0}$ has no component along ${\partial \over {\partial e}}$
one obtains by direct calculations:

$$V_{0}{_{_{\_\!\_\!\_\!\_\!}}{\!|}}
\Omega_{1}=-A_{i}P^{i}de + eF_{ik}P^{i}dX^{k}
\eqno (A.11).$$

\vskip 10pt

Further, tedious spinor algebra manipulations yield:

$$V_{1}{_{_{\_\!\_\!\_\!\_\!}}{\!|}}\Omega_{0}
= - Cde + e F_{ik}P^{k}dX^{i}
- e{F^{*}_{ik}}d(Y^{i}P^{k})=
- Cde + e F_{ik}P^{k}dX^{i}$$
$$ - d(e{F^{*}_{ik}}Y^{i}P^{k}) + eY^{i}P^{k}d{F^{*}_{ik}} +
(Y^{i}P^{k}{F^{*}_{ik}})de
\eqno (A.12).$$

\vskip 15pt

Using the fact that according to the
equations of motion the vector components of $V_{1}$ in the direction of
${\partial \over {\partial X^{i}}}$ and in the direction of
${\partial \over {\partial e}}$
are equal to zero one gets automatically:

$$V_{1}{_{_{\_\!\_\!\_\!\_\!}}{\!|}}\Omega_{1}= 0 \eqno (A.13).$$

\vskip 10pt

Putting together $(A.9)$, $(A.11)$, $(A.12)$, $(A.13)$ and
inserting $C=(Y^{i}P^{k}{F^{*}_{ik}}) - A_{i}P^{i}$ yields:

$$V{_{_{\_\!\_\!\_\!\_\!}}{\!|}}\Omega=
-d({1\over 2}m^{2} + s^{2} - {1 \over 4}e^{2})
- d(e{F^{*}_{ik}}Y^{i}P^{k}) + eY^{i}P^{k}d{F^{*}_{ik}}=
\eqno (A.14)$$

$$=-d({1\over 2}m^{2} + s^{2} - {1 \over 4}e^{2})
- d(e{F^{*}_{ik}}Y^{i}P^{k})=-dH \eqno (A.15)$$

\vskip 10pt

provided the last term in (A.14)
vanishes for all choices of $P_{i}$ and $Y_{i}$. That will always happen
if
the empty space  Maxwell's equations:

$$F^{*}_{[ik,n]}=0 \eqno (A.16)$$

\vskip 10pt

are fulfilled at the location
of the particle i.e. at its four-position in Minkowski space.

\vskip 10pt

This completes the proof of our assertion.

\vskip 10pt

Note that the first pair of Maxwell's equations is satisfied by virtue
of the fact that the external electromagnetic field is given by means of
a four-potential $A_{j}$ in the expression for the symplectic one-form
$\gamma$.  This automatically ensures that the symplectic structure
$\Omega$ is a closed two-form on T$\Delta$T.

\end{document}